Igor Timoshenko

Developing RFID library systems in the direction of integration into the global identification system EPC

*Abstract - The possibility of modification of the regulatory framework of RFID library systems in the direction of integration into the EPCglobal Network, based on existing harmonized standards of RFID technology. It is shown that this approach to the formation of the regulatory framework will improve the availability of RFID technology for libraries and contribute to the overall development of library technologies.*

The radio-frequency identification (RFID) technology has become ingrained in the technology of modern libraries. Comprehension of the gained experience and its consolidation in the form of generally accepted rules, whose implementation can provide its further development both in libraries and in other fields of activity, is a relevant task..

Broad introduction of the RFID technology in libraries occurred at the beginning of the 21st century. Libraries have accumulated a lot of experience, and it was reflected in the standards. In 2005, Denmark adopted the first National Standard, which defined the RFID tag application rules in libraries and was supported by many countries. In 2011, the ISO/TC46/SC4 Technical Committee adopted an International Standard ISO 28560 Information and documentation: Radio Frequency Identification in Libraries, which regulated the main technical parameters of RFID library systems, as well as the data exchange structures and protocols for automated library systems. Now the ISO 28560 standard is a group of standards, which consists of four parts. [1]. The standard was adopted in Russia. Today, all four parts of the standard exist in the Russian standardization system as identical to the international standard.

The introduction of this standard was an important step in the development of RFID library systems. At the same time, analyzing the content of the standard, it should be noted that its existing parts are not fully consistent with each other, which is a consequence of the historical situation of the use of RFID technology in libraries.

The first part of the standard described the data elements that can be stored in the tag memory and used in libraries.

In the second part, progressive ideas for encoding information were presented. In this part, it was declared that these ideas allow to create open library RFID systems, that would have a "direct relationship with supply chain activities" [1, part 2, ch. 4.2].

But the third part, based on the ideas of the Danish standard. It should be noted, that the principles of coding proposed in the third and second parts are mutually exclusive.

The fourth part appeared later, in 2014. In this part, was attempt to apply the ideas of the second part of the standard. The standard was developed for the use of UHF RFID equipment in libraries.

The presence of the third part of the standard is a statement of the fact that most of the libraries in the world using RFID have HF band equipment operating under the data model of the Danish standard or Similar. This situation is supported by the main manufacturers of specialized library equipment. Libraries using UHF RFID equipment are in a clear minority, "non-library" areas related to logistics. As one of the main reasons for this is the complete incompatibility of library RFID systems, UHF compared to HF systems, which occupies a stable position in libraries.

Solving these problems requires the efforts of software and hardware developers, which is associated with high financial costs. Such costs can be recouped by the widespread introduction of RFID technology into a large number of libraries. But the expansion of the market of RFID library systems today is limited by the relatively high cost of equipment available only to large libraries with good sources of Finance, as well as the shift of interest in the work of libraries with electronic resources. This is equally true for both HF and UHF systems.

This disappointing situation with the use of RFID in libraries exists against the backdrop of rapid and successful development of RFID technology in other areas, such as warehousing and transport logistics, which is close in functionality to library technology for working with hard-copy and paper-based documents. The way out of this situation is seen in the harmonization of the development of library technology with the General course of development of radio frequency identification technology.

Today, standard defines the operation of two independent library RFID systems. The common part for these systems - is only the first part. The difference in RFID equipment of different ranges (HF and UHF) is considered to be the origin of the systems incompatibility. But it is not true in everything. Especially if we consider the recent history of the development of RFID standards.

If we consider the development of the regulatory framework of RFID technology in the last ~ 15 years, we can see a clear trend of the transition of existing standardization centers from the statement of the current situation in the equipment market to the creation of a regulatory framework that determines the direction and stimulates the further development of RFID technology, that is, to create conditions for market development.

Let us illustrate this with examples from recent history. The first common RFID standards appeared in the late 90's - early 2000's in two standardization centers:

— Joint technical Committee ISO/IEC JTC1 "Information technology" /SC 31 "Automatic identification and data capture techniques" has developed a group of standards ISO/IEC 18000 for all types of RFID equipment, including HF and UHF.

— "EPCglobal" - an organization promoting the concept of" Electronic product code "as a single identifier for all RFID systems, and offering a standard for manufacturers of UHF equipment"EPC UHF Class 1 Generation 2 air interface specification"

It should be noted that the UHF group developed the idea of EPC - Electronic Product Code. This is a common technology for contactless identification of all objects.

The existence of two different groups of standards that determined the operation of similar types of devices and were not compatible with each other was a significant obstacle to the development of RFID systems. None of them was fully supported by equipment manufacturers. In two bands that are most used in practical applications, that is, HF (13.56 MHz) and UHF (850-960 MHz), leading manufacturers of HF RFID equipment (including the library type) began to use a standard ISO/IEC 18000-3 Mode 1 [2], while manufacturers of UHF use EPC C1g2 [3]. The use of equipment of a particular range in specific areas was determined by its characteristics and constraints that arise from the physical properties of electromagnetic waves. Along with this, logical units of the tags that operate in these bands differed significantly from each other, which, in addition to the difference in frequency ranges, made the HF and UHF RFID systems alternative to each other.

The first step towards harmonization of the two standardization trends was made by ISO/IEC JTC1. In 2006, a supplement was added to the existing ISO/IEC 18000 group of standards and an ISO/IEC 18000-63 [4] standard was adopted, which defines a data exchange protocol between UHF RFID devices, compatible with the protocol EPC C1g2.

The next step was the development in 2011 by the EPCglobal international organization together with GS1 of a standard EPC Class 1 [5], which defined EPC protocol concepts for HF equipment. The new standard was supported by ISO/IEC JTC1/SC31 by adopting a similar supplement Mode 3 to the ISO/IEC 18000-3 standard.

The emergence of common standard approaches to the RFID equipment production and application in the most popular frequency bands has created a fundamental opportunity to implement the initial EPCglobal concept on the use of a single electronic product code EPC for the identification of accounting items in RFID systems of various specializations, including library ones.

The ability to «transparently» work EPC RFID system in two ranges, along with the use of EPC tags, requires the improvement of double-frequency RFID tag readers. The creation of such tag readers is a highly technical task. A tentative move in this direction was made by FEIG Electronic company, which began the production of ID ISC.PRHD102 mobile readers in 2013, which supported simultaneous operation in the HF/UHF bands.

Commercial availability of HF EPC tags occurred in 2013 when NXP Semiconductors company launched ICODE ILT chips that complied with ISO/IEC 18000-3 Mode 3 (EPC Class 1 HF). Based on these chips, it is possible to produce library HF EPC tags.

In library standard ISO 28560, the use of equipment based on EPC system is determined by the fourth part. The data structures presented in the standard are oriented to RFID tags having block memory organization defined in the standard "EPCglobal" as "Class 1 Generation 2" (EPC C1g2). The standard is focused on UHF tags, which formally narrows its scope, because in fact, the memory structure described in the standard currently can RFID tags of two types :

— HF - ISO/IEC 18000-3 Mode3 (EPC Class 1 HF).
— UHF - ISO/IEC 18000-6 type C (EPC C1g2).

The standard can be assigned to both types of labels equally, so it will be productive to point out in the standard not on the frequency range, but on the use of EPC-oriented equipment system. With this approach, ISO 28560-4 can be considered an important step in the development of RFID library systems, and will fit into the General direction of the development of RFID technology.

The first and second parts of the ISO 28560 standard describe data elements and how to encode them to store labels in memory. At principle level, this can be implemented in any label of any frequency band with sufficient memory. The fourth part of the standard, based on its title, refers to labels that have a block memory organization. The "block memory organization" feature is not associated with the frequency range, but it is associated with the adaptation of equipment for work in EPC information systems. Specific examples of such equipment, at a regulatory framework level, are given in the section "Normative references". In this section, it is appropriate to specify all label types with the appropriate memory organization.

Another question proposed for discussion is the principle of forming a Unique Item Identifier (UII), placed in the block 01 memory labels (EPC-block).

The EPC concept is being actively developed and reflects the overall development of the information environment and global information systems such as the EPCglobal Network and IoT. Items used in libraries (library documents) enter libraries from the outside world, and after becoming documents of the library holdings, can also leave the libraries and be in the outside world, while participating in the technological processes of the library. The possibility of including library documents into the external information systems processes, as well as using the capabilities of external information systems for libraries, can significantly enrich and give an additional impetus to the development of library technologies, as well as reduce the cost of the use of RFID technology in the libraries themselves. To do so, library documents must be identified, along with ILS, by external information systems. When used in libraries EPC RFID equipment, it is logical to use the format of the UII code compatible with the format of the EPC code.

In connection with the above, it should be noted that the data format proposed in the standard is incompatible with the principle of code generation in EPC systems. The absence of such compatibility makes RFID library systems local and limits the scope of their application libraries only. This hinders the integration of library technology into related radio frequency identification technology areas. In principle, this possibility exists based on the technical characteristics of the labels used in library RFID systems and technological peculiarities of cataloging documents in the library.

The Electronic Product Code (EPC) offered by EPCglobal/GS1 is a numeric identifier that is unique to each accountable material item. Currently, the most wide spread codes have lengths of 64 and 96 bits. There are standard and are planned to be implemented as 198 bits code. The total code length determines the possible data field length, and as a consequence, the width of the code space and the freedom to choose the data representation format. The basic structure of the EPC code consists of four fields, whose intention is specified as follows:

— the header specifies the code type;
— the number of the supplier/owner determines the organization – owner of the accounting item;
— the item class determines the typical identity of the accounting item;
— the serial number identifies the definite item copy.

The data fields are represented in the code as a rigid fixed-length fields structure, whose size is determined by the code type. The values of the first two fields are regulated and assigned by the EPCglobal/GS1 standards. The last two fields are assigned by the owner organization on the basis of local technological requirements. These fields can be rerecorded when passing from one stage to another.

When a document enters a library upon being tagged at the early stages of the supply chain in accordance with ISO 28560-4, the entire block 01 (EPC) of its tag memory should be represented by a unique identifier of the accounting item (UII) and an AFI byte. This leads to data loss on the EPC code and the format of the recorded information does not correspond to the EPC format; thus, the tag is no longer recognized by EPC RFID systems. If the tag is returned to the automated system, for example, being delivered to a receiver through a transport company, postal service, or being offered for sale, it is necessary to restore the EPC code in its memory with a "library" data loss. It is technically possible to use the EPC identification code in ILS, but at present this is not provided for by library standards.

As an example, illustrating the possibility of integrating RFID systems, it is possible to consider the marking of printed publications in their manufacture in the printing house. The publisher can assign its values to the type of publication and serial number at the initial stage of the logistics chain. When such documents enter a library, these fields can be reassigned in accordance with the library cataloging rules and the value of the provider/owner number field can be used for cataloging as the publisher's or provider's organization identifier. The "object class" field can be used to store the publication type UNIMARC or ONIX identifier. In this case, the field "serial number" only is subject to rerecording, which contains from 36 to 180 assigned bits for different types of EPC tags. This field may contain data elements that constitute the unique identifier of the item (UII) in accordance with ISO 28560-4.

In the calculation, the total length of the UII, according to the Danish data model, is 19 bytes, which consists of the following components:

— primary identifier – 16 bytes;
— ISIL code – 11 bytes;
— set information – 2 bytes

For encoding by URNCode40 rules, the total code length is 12 bytes. Thus, the total length of the entry in the "serial number" field together with the added value of the AFI byte is 104 bits. The resulting field size does not exceed the maximum possible size for the EPC format for SGTIN198 (140 bits) and is fully contained in the complete EPC code in the 01 memory block of NXP ICODE ILT labels, which size is 240 bits. The ISIL code (11 bytes) can be placed in memory block 11 (user memory).

The above example shows only a fundamental possibility of integration standard library automation systems and RFID systems based the EPC standards. This possibility now exists and follows from the general logic of the development of information systems, in particular, RFID technology. The libraries will win from the integration of library RFID systems with EPCglobal Network. It will enable to further develop the RFID systems to achieve the repeated use of RFID tags, starting from publishing house. This will enable to reduce the cost of implementing and using RFID technology in libraries. In addition, it will allow to participate in the development the "IoT" together with EPC systems.

Today, the concept of IoT (Internet of Things) is actively developing. Such development involves the emergence of diverse "transparent" functionally localized information systems. These systems are based on standard communications and standard identifiable objects. Library objects can participate in these systems. This can greatly expand the functionality of ILS. The IoT systems with library functionality can partially replace the functions of specialized library automation systems. Such development of RFID systems will promote the development of library technologies.

The inclusion of library tagged documents in the EPCglobal Network can significantly increase their mobility in the delivery services in the framework of the interlending system in the future, as well as the availability of the active stock for readers through the extensive use of new global identification technologies. In general, it can be argued that the use of the RFID technology in libraries, based on a regulatory system that is harmonized with global identification technologies, will significantly enhance the integration of a conventional library collection along with electronic documents into the modern information space within the framework of new concepts for the development of library technologies.

**Information about the author:**

Igor Timoshenko - PhD in engineering, leading researcher, Chief AS Technologist, Russian National Public Library for Science and Technology, Chairman of GOST RU TC355/SC4/WG1 "RFID in Libraries"